\documentclass[11pt]{article}
\usepackage[margin=1in]{geometry}
\usepackage{amsmath,amssymb,bm}
\usepackage{graphicx}
\usepackage{booktabs}
\usepackage{xcolor}
\usepackage{microtype}
\usepackage[super,sort&compress]{natbib}
\setcitestyle{super,comma,open={},close={}}
\usepackage[hidelinks]{hyperref}

\title{Inverse Designing Reconfigurable Piezoelectric Tensors via Architectural and Modal Programming}

\author{%
  \parbox{0.9\textwidth}{%
    \centering
    Anish Satpati\textsuperscript{1},
    Marco Maurizi\textsuperscript{2},
    Hyunmoon Cho\textsuperscript{1},
    Victor Couedel\textsuperscript{1},
    Nicolas Mule\textsuperscript{1},
    Xiaoyu (Rayne) Zheng\textsuperscript{1,*}\\[1em]
    {\small
    \textsuperscript{1}Department of Materials Science and Engineering,
    University of California Berkeley, Berkeley, 94720, USA\\[0.5em]
    \textsuperscript{2}The Italian Institute of Artificial Intelligence (AI4I),
    Turin, Italy\\[1em]
    \textsuperscript{*}Corresponding author.
    Email: \texttt{rayne23@berkeley.edu}
    }
  }%
}
\date{}

\begin{document}
\maketitle

\section*{Abstract}

Piezoelectricity arises from the coupling between electric polarization and mechanical deformation, described by a third-rank tensor whose coefficients are constrained by crystal symmetry. In natural crystals, atomic periodicity and energetic stability restrict accessible symmetries, limited to 20 distinct anisotropies that can exist. Here we establish a general framework, based on graph representation, for controlling the full effective piezoelectric tensor through architecture by resolving how macroscopic symmetry, local polarization and deformation of structural elements act together. This enables each coefficient to be traced to the members that reinforce or cancel it and be controlled through targeted geometric changes. We further uncover frequency-selective tensor reconfiguration, in which excitation frequency switches the dominant piezoelectric coefficients of a single architecture. These principles open a broader class of applications in which a single piezoelectric architecture can deliver programmable sensing, actuation and mode conversion across multiple directions and deformation states. We realize a multi-degree-of-freedom actuator that integrates distinct actuation modes in one block and reduces volume by 90\% relative to conventional multi-stack assemblies.

\section*{Main}

Piezoelectricity enables reciprocal conversion between electrical and mechanical energy, underpinning precision actuators, ultrasonic transducers, energy harvesters, implantable sensors and microelectromechanical systems \cite{Wang2025Nanopositioner,Hu2023CardiacUltrasound,Xu2010Nanowire,Zhou2026Implantable,DePonti2024PiezoMEMS}. The coupling is governed by a third-rank piezoelectric tensor \(d_{ijk}\), which relates an electric field applied along direction (k) to the mechanical strain generated in directions (i) and (j), or conversely relates an applied stress to the induced electric polarization. In natural crystals, however, these coefficients cannot be chosen independently. Crystal symmetry determines which tensor components are allowed, which must vanish, and which are related. Only 20 of the 32 crystallographic point groups are piezoelectric, and commonly used ferroelectrics therefore access only a small subset of the full tensor space \cite{Nye1985,deJong2015}. For example, tetragonal ferroelectrics have only three independent piezoelectric coefficients \cite{Nye1985}. Fully populated piezoelectric tensors require very low crystal symmetry, but such ferroelectric phases are rare because they are difficult to stabilize and often occur only under narrow compositional, field, or strain conditions \cite{VanderbiltCohen2001,Noheda1999,Chen2011,Paull2022}. Commercial piezoelectric sensors and actuators use piezoelectric stacks whose response is along fixed material directions, limiting the motion generated by each element \cite{Crawley1987}. Generating multi-axis motion therefore requires several piezoelectric elements and flexure mechanisms to redirect and amplify their displacement, substantially increasing device footprint, and mechanical complexity \cite{Yong2012,LiXu2011}.

Additive manufacturing has enabled new opportunities to tune effective piezoelectric coefficients by architecting the macroscopic geometry \cite{Bertoldi2017,Kadic2019,Zok2016,Zheng2014,Kim2014,Cui2019,Cui2022,ShiAkbarzadeh2019,Shi2024,Tao2022,Ammu2024}. These architected metamaterials are 3D architectures made of microscopic, piezoactive ligaments which once poled, can exhibit unique effective piezoelectric property absent from base material \cite{Yang2019,Yu2023,Shi2025,Sharma2025}, such as electro-auxetic effect, amplification, bending and twist effect for a broad array of sensors and actuator applications. However, designing these coefficients remains a significant challenge because each effective coefficient combines macroscopic symmetry, local poling and deformation across the connected ligaments. Changing one ligament redistributes both the electric field and mechanical deformation throughout the network making the design of arbitrary piezoelectric tensor elusive. Additionally, the static tensor alone is insufficient for ultrasonic transducers and resonant actuators operating under alternating electric fields \cite{Hagood1991,Li2022Modes,Ogi2004}. This is because the same geometry also determines the structure's resonant vibration patterns \cite{Airoldi2011,Casadei2012}, causing normal and shear responses to fluctuate with frequency. Predicting and controlling this frequency dependence could enable multimodal piezoelectric devices that perform distinct sensing or actuation functions at different operating frequencies.

Addressing both the static and frequency-dependent problems requires a representation that follows how electrical and mechanical effects pass through each geometric element. A graph is a mathematical representation of a system in terms of discrete elements and the connections between them, providing a compact description of complex networks and their topology. Graph representations have been widely used across disciplines including chemistry, biology, and mechanical networks to describe connected systems \cite{XieGrossman2018,Reiser2022,Merchant2023,Maurizi2025}. Here, we introduce a graph-based framework for piezoelectric architected materials, representing the 3D architecture as a graph in which junctions are nodes and ligaments are edges. By assigning each edge to a physical ligament, the framework directly links local structural elements to the macroscopic piezoelectric tensor and provides a basis for identifying how architecture controls its magnitude, sign and frequency-dependent response. We establish a general route to inverse design the full piezoelectric tensor through 3D architecture, enabling extreme anisotropy, strut-level control of individual coefficients, and frequency-selective switching different coefficient tensor components. This transforms a single piezoelectric material into a reconfigurable, multifunctional platform and enables compact devices with capabilities inaccessible to conventional piezoelectric materials.

\section*{Results}

\subsection*{Architecture releases crystallographic tensor constraints}

\begin{figure}[p]
\centering
\includegraphics[width=\linewidth,height=0.60\textheight,keepaspectratio]{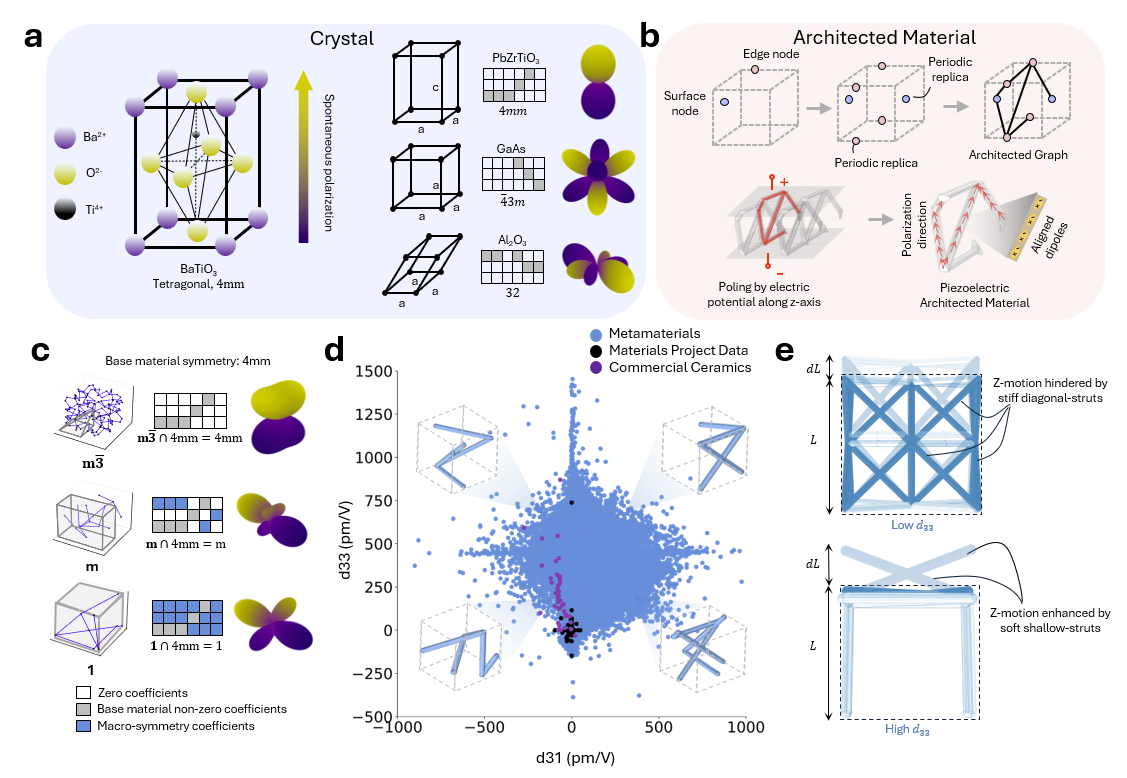}
\caption{\textbf{Architecture lifts crystallographic constraints on the piezoelectric tensor.} \textbf{a}, Representative crystal classes and their symmetry-constrained piezoelectric tensors. \textbf{b}, Generation of periodic unit-cell graphs and alignment of the local polarization with the truss members. \textbf{c}, Architectural symmetry breaking produces fully populated effective tensors.  \textbf{d}, Piezoelectric metamaterials expand the property space beyond Materials Project crystals and commercial ceramics. \textbf{e}, In the octet truss, diagonal struts resist changes in height and limit \(d_{33}\). Shallow struts in the graph architecture allow greater \(z\)-directed motion via mixed-mode deformation.}
\label{fig:design-space}
\end{figure}

Crystallographic point-group symmetry confines natural piezoelectric materials to discrete tensor forms (Fig.~\ref{fig:design-space}a). For example, tetragonal \(4mm\) BaTiO\(_3\) has five non-zero matrix entries defined by three independent coefficients, whereas cubic \(\bar{4}3m\) GaAs and trigonal \(32\) \(\alpha\)-quartz have one and two independent coefficients, respectively \cite{Nye1985,deJong2015}. Here, we start from a single \(4mm\) PZT constituent material and transfer tensor design from chemistry to architecture. We encode the piezoelectric material as a graph \(G=(V,E)\), in which nodes \(V\) denote junctions and edges \(E\) denote struts in the periodic 3D representation of the truss architecture. We start by establishing elementary units from nodes randomly placed on the boundaries of a cube (Fig.~\ref{fig:design-space}b). We construct each unit cell from scratch by randomly placing face and edge nodes, with periodic copies forming six- and eight-node configurations that span a broad range of geometries. Starting with this simple and unconstrained assembly enables a substantially wider space of architectures and forms the first basis for controlling the complete piezoelectric tensor through geometry. Once the node positions are defined, we generate sequences of N-2 node labels that connect unique sets of N-1 struts, producing graphs with diverse topologies (Fig.~\ref{fig:design-space}b; Supplementary Section~1; Supplementary Fig.~S1). The poling of the architecture enables the dipoles to be aligned within each strut comprising the unit cell. Therefore, the local electric field within a 3D architecture determines the shape and magnitude of the tensor, with its direction setting the local polarization axis and its strength determining the degree of domain alignment (i.e., magnitude of coefficient) \cite{Shi2025} (Methods, Supplementary Sections~2.3; Supplementary Fig.~S3). Individual struts can consequently reinforce, oppose or redirect one another, allowing the same piezoelectric material to exhibit a wide range of tensor coefficients, going beyond solid constituent material.

The graph design space sets the positions and connections of the struts, while the local poling field sets the direction and magnitude of their responses. The remaining design variable is the macroscopic symmetry of the truss structure which determines the effective tensor coefficients that are permitted. The tensor anisotropy of the architected material can be determined by the symmetry operations shared by the base material and that of the poled architecture, as \(G_{eff}=G_{graph}\cap G_{base,z}\) where \(G_{graph}\) is the point group of the architected material and \(G_{base,z}\) is the symmetry of base material poled along the global z-direction \cite{Nye1985,deJong2015}. (Supplementary Section~3; Supplementary Fig.~S4). Here, a structure that shares the same symmetry or contains all symmetry operations present in \(G_{base,z}\), possesses the same form of tensor (i.e., same location of non-zero coefficients, but can vary in magnitude). To establish new tensor shape and activate other non-zero coefficients, operations such as mirror planes and rotation axes removal allow us to change the symmetry from \(G_{base,z}\), activating new non-zero coefficients. To independently tune and activate all 18 coefficients, we reduce all symmetry elements of the graph to point group 1, activating all 18 entries of the \(3 \times 6\) tensor (Fig.~\ref{fig:design-space}c) \cite{Yang2019,Yu2023}. Together, generating graphs that incorporate local electric fields, and macroscopic symmetry produces a substantially broader range of coefficient magnitudes and tensor anisotropies (Fig.~\ref{fig:design-space}d) than the inorganic crystal tensors computed from first principles in the Materials Project database \cite{deJong2015}. We therefore complement the crystal benchmark with experimentally measured piezoceramics, which our graph architecture also exceeds (Fig.~\ref{fig:design-space}d) \cite{Liu2018PiezoReview,Tichy2010Sensorics,Rodel2009LeadFreeReview}. Further comparisons of tensor-space coverage and coefficient magnitude are provided in Supplementary Section~1.1.

Beyond activating all 18 piezoelectric coefficients and spanning a broad anisotropy space, our unconstrained graph construction also enables individual coefficients to reach magnitudes far beyond those accessible in conventional architectures. The high-\(d_{33}\) designs reveal that these extreme responses arise from programming anisotropic compliance through architected load paths (Fig.~\ref{fig:design-space}e). This enhancement arises from the distinct deformation modes supported by the architecture: axial stretching of struts strongly resists transverse motion, whereas bending, rotation, and mechanism-like deformation provide a compliant pathway along the actuation direction. The resulting combination of high lateral stiffness and low (z)-direction stiffness suppresses displacement in (x) and (y) and concentrates the piezoelectric deformation along (z). This enables a \(d_{33}\) \(\sim\) 3 times larger than previously studied conventional architectures like the octet-truss. The enhanced \(d_{33}\) therefore originates from architecture-enabled redistribution of deformation, rather than from any change in the intrinsic properties of the constituent material (Supplementary Section~4).

\subsection*{Inverse design of prescribed and extreme piezoelectric tensors}

Here, by leveraging the graph representation, we establish the inverse design framework to obtain architectures with prescribed tensor anisotropy, extreme individual coefficients or specialized deformation patterns by activating all 18 coefficients. To learn the architecture--property relation, we used coupled electro-mechanical piezoelectric beam simulations to calculate all 18 homogenized coefficients for 611,239 periodic trusses and trained a graph neural network (GNN) to predict the complete tensor from each structure's node positions, connectivity and strut geometry (Methods, Supplementary Section~5; Supplementary Fig.~S5). Through message passing, the GNN exchanges node and strut features along each edge and aggregates them at connected junctions, progressively capturing how local interactions determine the homogenized tensor. The GNN successfully reproduces the signs, relative magnitudes and anisotropy of the complete tensor, allowing us to define the following inverse-design strategy. For a user-defined tensor anisotropy, the model ranks the graphs in the training set and selects the \(k\) closest matches as initial candidates. Starting from each candidate, the optimizer varies the free nodal coordinates while preserving connectivity and periodicity. At every iteration, it updates the strut lengths, recalculates the radii to maintain constant relative density, rotates the local piezoelectric tensors to the new orientations and predicts the resulting effective tensor. The Frobenius distance between the predicted and target tensor is minimized by differentiating the predicted response with respect to the nodal coordinates. This procedure ensures graph connectivity with continuous refinement of geometry while keeping every design variable directly associated with the unit-cell architecture (Supplementary Section~6; Supplementary Fig.~S6).

\begin{figure}[p]
\centering
\includegraphics[width=\linewidth,height=0.60\textheight,keepaspectratio]{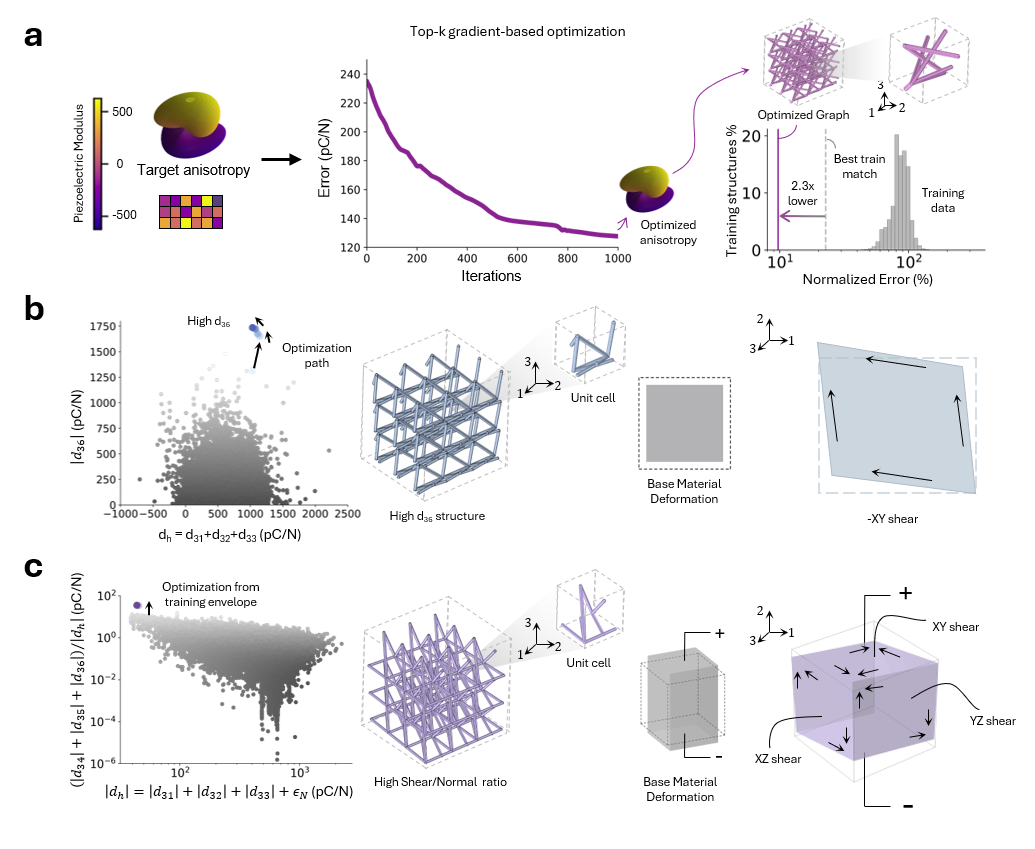}
\caption{\textbf{Inverse design of arbitrary anisotropy and extreme piezoelectricity.} \textbf{a}, Starting from the closest training structure, top-\(k\) optimization yields an optimized graph with a 2.3-fold lower normalized tensor error when designing for arbitrary tensor anisotropy. \textbf{b}, Maximizing \(|d_{36}|\) moves the design beyond the training-data envelope, resulting in a 32\% improvement relative to the starting structure. The optimized lattice produces \(xy\) shear, unlike the base-material deformation. \textbf{c}, Maximizing the shear-to-normal ratio \((|d_{34}|+|d_{35}|+|d_{36}|)/(|d_{31}|+|d_{32}|+|d_{33}|+\epsilon_N)\) produces a structure with strong shear response and suppressed normal deformation. The term \(\epsilon_N\) accounts for the GNN prediction error.}
\label{fig:inverse-design}
\end{figure}

We demonstrate the inverse design of piezoelectric behaviors include full-tensor anisotropy generation, coefficient-specific enhancement and tailored combinations of deformation modes. We first demonstrate full-tensor matching by prescribing all 18 coefficients of a target response with user-defined longitudinal anisotropy (Fig.~\ref{fig:inverse-design}a). The optimizer identifies a new geometry that reproduces the prescribed tensor more closely than any structure in the training set, demonstrating inverse design beyond retrieval (Fig.~\ref{fig:inverse-design}a).

Our framework can also selectively amplify individual piezoelectric coefficients beyond the sampled property range. For example, we target \(d_{36}\), which couples an electric field \(E_z\) to \(xy\) shear (Fig.~\ref{fig:inverse-design}b) \cite{MiaoLi2015,Yang2019,Yu2023}. Optimization increases this coefficient from \(1315.1\) to \(1735.8~\mathrm{pC\,N^{-1}}\), a \(32\%\) improvement that exceeds all structures in the training set. The resulting deformation exhibits the corresponding shear mode, confirming that the enhanced tensor coefficient produces the intended electromechanical performance.

Beyond individual coefficients, prescribed deformation generated from a combination of piezoelectric coefficient while others being suppressed can be generated. As a representative case, we set the objective to favor large pure shear coefficients \(d_{34},d_{35},d_{36}\) under \(E_z\), while the normal coefficients \(d_{31},d_{32},d_{33}\) being suppressed to zero. In Fig.~\ref{fig:inverse-design}c, the optimized structure amplifies selected shear components while suppressing normal deformation, producing a predominantly shear-driven motion directly through the effective properties of the architected material. Together, these examples show that inverse design can prescribe full-tensor behavior, create otherwise inaccessible electromechanical couplings, and program specialized deformation modes through geometry. The same framework could target other application-specific motions, such as pure angular actuation in torsional micromirrors while suppressing unwanted translation \cite{Grinberg2017Twisting,Zhu2011Micromirror}. Independent beam-model validation of all three optimized designs is provided in Supplementary Section 6.3.

\subsection*{Reconfigurable piezoelectric tensor anisotropy}

Most piezoelectric devices operate under dynamic electrical excitation, yet their behavior is still commonly designed using a quasi-static piezoelectric tensor measured below structural resonance \cite{IEEE1761987,Barzegar2004}. In piezoelectric structures, structural vibration modes can generate distinct strain fields and activate different electromechanical couplings, causing the effective tensor to change substantially with frequency. This mode-dependent tensor evolution remains poorly understood and is rarely controlled in device design \cite{Hagood1991,Li2022Modes,Liu2000,Hussein2014,Srivastava2015,Airoldi2011,Casadei2012}. Here, we uncover how individual eigenmodes reshape all 18 components of the effective piezoelectric tensor and demonstrate a new capability to reconfigure the electromechanical response through frequency selection. By changing the excitation frequency alone, the same architecture can selectively activate different tensor components and deformation modes without changing its material or geometry (Methods, Supplementary Section 7).

\begin{figure}[p]
\centering
\includegraphics[width=\linewidth,height=0.60\textheight,keepaspectratio]{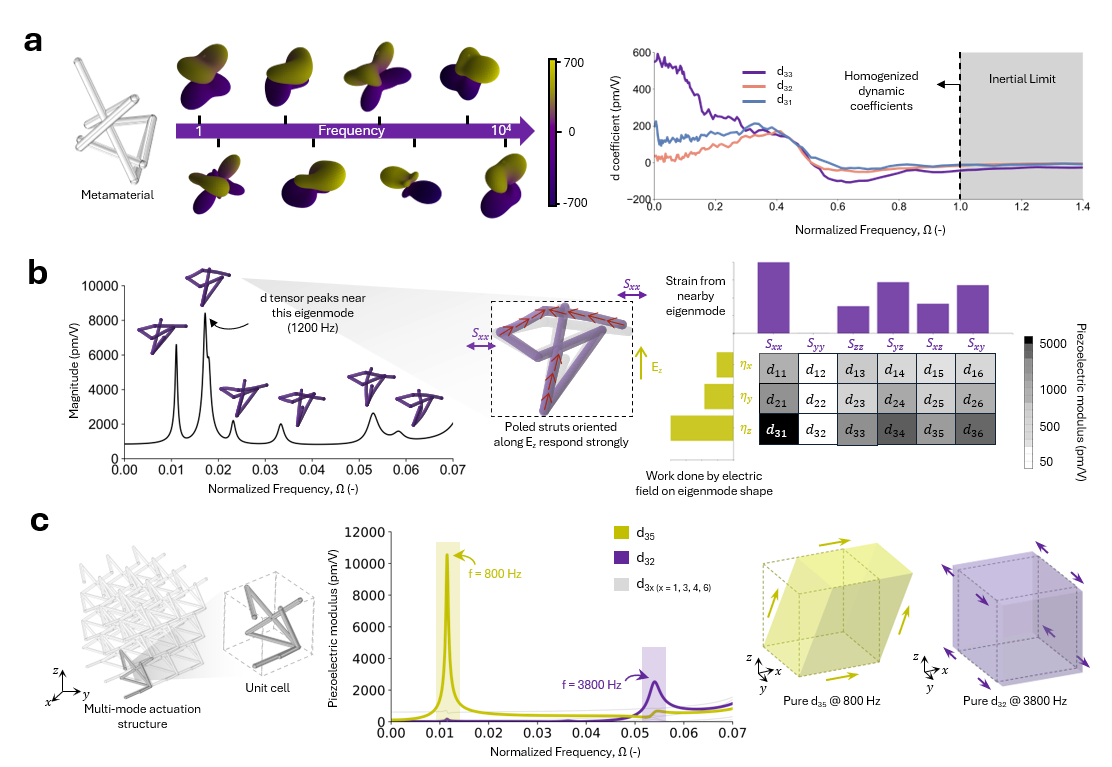}
\caption{\textbf{Frequency-domain control of piezoelectric tensor.} \textbf{a}, Evolution of the piezoelectric anisotropy of a representative uniformly poled truss with frequency. Representative homogenized coefficients \(d_{31}\), \(d_{32}\), and \(d_{33}\) vary strongly with normalized frequency \(\Omega\), before going to zero at the inertial limit denoted by the shaded region beyond \(\Omega = 1\). \textbf{b}, For a representative structure, the dynamic tensor magnitude peaks close to a structural eigenfrequency at 1200 Hz (0.017 normalized frequency), with the deformed shapes shown in the inset at each eigenfrequency. The eigenmode corresponding to the global peak produces strong \(S_{xx}\) strain and is strongly driven by \(E_z\), making \(d_{31}\) dominant in the complete frequency-dependent \(3\times6\) tensor shown in the heat map. \textbf{c}, Mode-selective activation enables different coefficients within a single architecture. Distinct resonances couple the same poled topology to different deformation channels, producing an approximately pure \(d_{35}\) response near 800 Hz and an approximately pure \(d_{32}\) response near 3800 Hz, while the remaining \(d_{3J}\) coefficients remain comparatively weak.}
\label{fig:frequency}
\end{figure}

The resonances described here arise from the distribution of mass and stiffness created by the architecture. The truss structure introduces internal motion at the unit-cell scale because its struts can stretch, bend and rotate in different combinations. Changing the graph connectivity or node positions therefore controls the mode shapes, resonance frequencies and associated strain patterns. We combine this architectural freedom with frequency selection to activate different tensor components and modify the deformation mode in a single structure. Figure~\ref{fig:frequency}a shows that near resonance, the individual coefficients increase, decrease or reverse sign differently because each frequency excites a different combination of normal and shear deformation. Because the relevant frequency scale depends on the unit-cell size \(a\) and constituent wave speed \(c\), we report \(\Omega=fa/c=\omega a/(2\pi c)\) to compare responses across materials and cell sizes (Fig.~\ref{fig:frequency}a; Supplementary Section~7.2) \cite{Trainiti2016,Lucklum2017}. Under small deformation, the dynamic response is the sum of the structure's mechanical eigenmodes. The graph connectivity and node positions determine the modal strain, while the orientation and local polarization of the struts determine the work done on that mode by the applied electric field. Their combination determines which coefficients are expressed at a given frequency, and a tensor coefficient is strongly amplified only when the mode produces substantial strain and couples strongly to the electric-field drive (Supplementary Section~7.3).

Modal strain and electric-field drive can be calculated from each eigenmode to predict which tensor coefficients are likely to dominate near resonance, without running a full frequency-domain simulation. For the structure in Fig.~\ref{fig:frequency}b, the resonant mode produces predominantly \(S_{xx}\) strain and is driven most strongly by \(E_z\), predicting a dominant \(d_{31}\) response that the finite-element frequency sweep confirms. Using modal strain and electric-field drive as selection criteria, we identify an architecture that reconfigures normal and shear actuation with frequency (Fig.~\ref{fig:frequency}c). At \(800~\mathrm{Hz}\), coordinated strut motion produces pure \(xz\) shear with little normal extension, which gets switched at \(3800~\mathrm{Hz}\), to a pure \(y\)-direction actuation where the struts collectively change the structure's height with little shear. The same \(E_z\) input can therefore access two mechanical outputs encoded by the architecture. Unlike approaches based on external shunt circuits or separately addressed piezoelectric units, this response is selected directly from the intrinsic mode spectrum \cite{Hagood1991,Li2022Modes,Airoldi2011,Casadei2012}. Both responses and the modal analysis are validated in Supplementary Sections~7.4.

\subsection*{Architecture-programmed motion for nanopositioning}

Controlling a robotic fingertip to position and orient an object in 3D requires six degrees of freedom (DOF), comprising translation along three axes and rotation about each axis \cite{Zhang2023}. Conventional systems generate these motions by assembling multiple uniaxial piezoelectric stacks with flexure mechanisms that redirect and combine their displacements, resulting in bulky and mechanically complex structures \cite{Wang2025Nanopositioner}. Here, we achieve these complex motion functions directly from compact piezoelectric metamaterial building blocks whose deformation modes are encoded by architecture and programmed electrically. The inverse-design framework decomposes the desired six-degree-of-freedom motion into elementary orthogonal strain modes, then designs metamaterial elements that directly produce each required normal or shear response (Fig. 4a). It next determines the minimum number and arrangement of these elements needed to reconstruct the target motion. By programming the voltage amplitude, polarity and phase applied to the individual metamaterials, translations, rotations and compound motions can be generated from the metamaterial arrays. This converts a mechanically complex motion-control problem into programmable elementary material responses.

\begin{figure}[p]
\centering
\includegraphics[width=\linewidth,height=0.60\textheight,keepaspectratio]{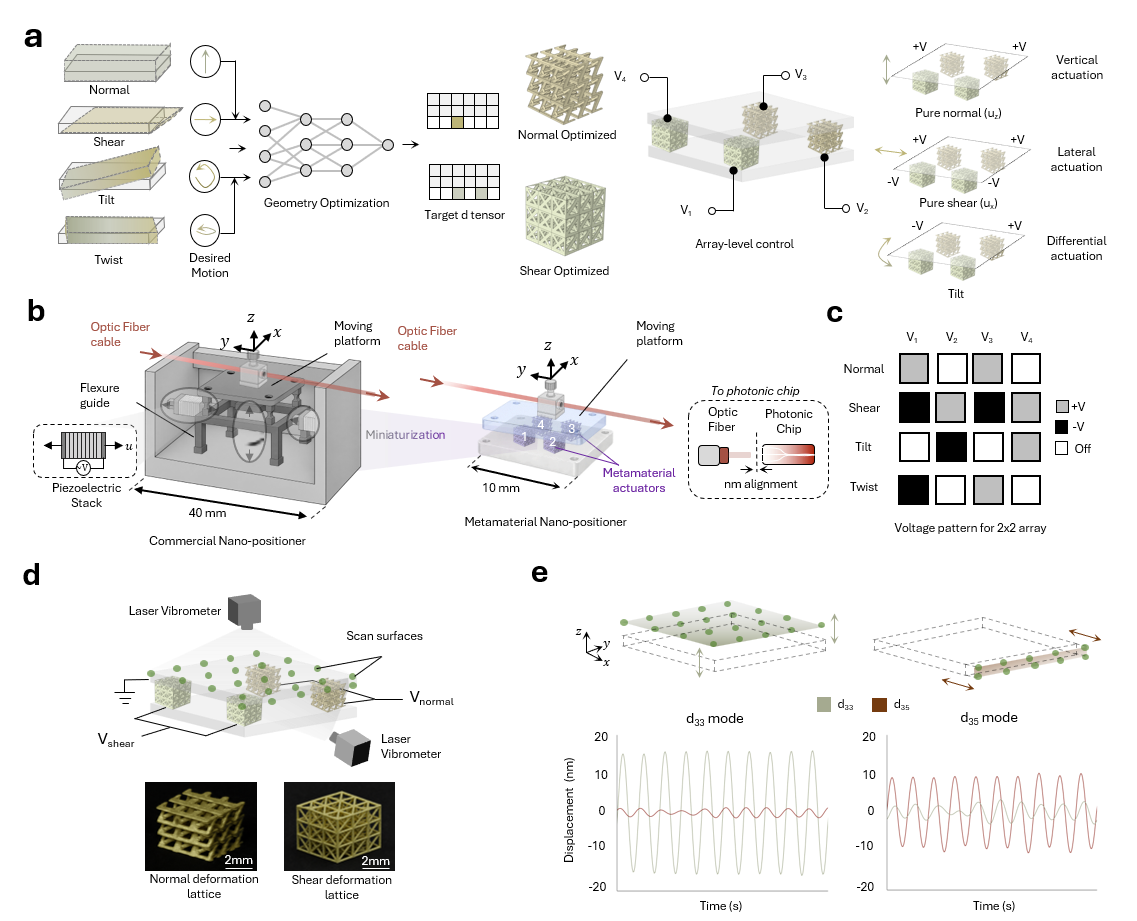}
\caption{\textbf{Geometry-programmed nanopositioning.} \textbf{a}, Desired normal, shear, tilt and twist motions are mapped through inverse design to obtain normal- or shear-optimized truss architectures. Voltage patterns then combine or cancel their displacement components to produce a variety of motions. \textbf{b}, Commercial fibre-to-chip nanopositioners occupy large footprint due to several piezoelectric stacks and nested flexure guides whereas the smaller metamaterial platform generates normal and lateral motion directly through its architected elements. \textbf{c}, Voltage patterns applied to the four elements of the \(2\times2\) metamaterial array to generate normal motion, shear, tilt and twist. \textbf{d}, Laser Doppler vibrometry configuration used to measure the normal and lateral displacements of the assembled platform. Fabricated normal- and shear-response lattices are shown below (scale bar, 2 mm). \textbf{e}, Reconstructed platform deformation and measured displacement shows the de-coupling ability to produce normal and shear deformations on the platform.}
\label{fig:nanopositioner}
\end{figure}

We demonstrate this concept using nanopositioning of an optical fiber relative to a photonic chip, a critical alignment task in modern photonic packaging where precise vertical and lateral positioning with at least four degrees of freedom (namely, vertical normal motion, lateral shear, tilt and twisting motion) is required to maximize optical coupling after coarse assembly   (Fig.~\ref{fig:nanopositioner}b). Conventional nanopositioning stages generate these motions using multiple uniaxial piezoelectric stacks together with flexure mechanisms that redirect and combine their displacements into different motion directions. This approach requires separate actuator and motion-conversion components, resulting in bulky and mechanically complex assemblies. In contrast, our inverse-design framework breaks down the required motions into two basic responses to an applied electric field \(E_z\). This requires one metamaterial design dominated by \(d_{33}\), which extends or contracts along z, and another with coupled \(d_{33}\)–\(d_{35}\) response, which combines the normal deformation in z with shear in the xz plane. These responses are sufficient because normal and shear deformation provide vertical and lateral motion, while arranging two elements of each architecture in a 2×2 array with individual voltage control enables tilt and twist degrees of freedom, achieving all four required motions (Fig.~\ref{fig:nanopositioner}c). The resulting actuator measures approximately \(10\times10\times5~\mathrm{mm^3}\), compared with the \(40\times40\times40~\mathrm{mm^3}\) representative commercial stage \cite{PINanoCube2016} (Supplementary Section 9.4).

We experimentally realize this compact motion-control concept using a four-element PZT metamaterial array. Two elements provide predominantly normal deformation, while two provide coupled normal and shear deformation, giving the elementary responses required for vertical and lateral positioning. The measured deformation modes agree with the designed responses (Fig. 4d). After assembly beneath a common platform, electrical control of the four elements generates either normal motion or lateral motion while suppressing the unwanted normal component. The resulting platform therefore reconstructs the required positioning motions directly from architected material responses, experimentally demonstrating compact multi-axis actuation without a separate mechanical motion-conversion mechanism (Fig. 4e).

\section*{Outlook}

The effective piezoelectric tensor of a metamaterial arises from the combined action of macroscopic symmetry, local poling and structural deformation. Symmetry determines the permitted responses, the local field sets each strut's polarization, and the connected structure determines how these contributions add or cancel. The graph representation links this collective response to individual geometric struts. The large-\(d_{33}\) structures further show that strong coupling requires local piezoelectric strain to align with a compliant macroscopic deformation. Their response combines axial strut actuation with bending and joint rotation, establishing a mixed-mode deformation for high performance. Frequency provides another dimension of control, allowing the same architecture to produce different tensor responses under harmonic excitation. The coupling between the electrical excitation and each mode’s strain pattern helps determine the dominant coefficients. This mechanism could enable multi-axis actuators with fewer electrical connections, frequency-selective sensors, and transducers that switch functions without additional electrodes or moving parts. The printed nanopositioner demonstrates how PZT elements with different architectures can be combined and electrically controlled to achieve all required degrees of freedom without separate flexures or motion-conversion mechanisms, reducing device volume by \(90\%\). Together, the design framework and frequency based mode-selection offer a route to piezoelectric devices for precision positioning, microrobotics and adaptive sensing, with architecture defining the available responses and electrical excitation controlling how they are combined or selected during operation.

\section*{Materials and Methods}

\subsection*{Graph construction and numerical poling}

We generated periodic unit cells from 13 arrangements of paired face and edge nodes: four arrangements with six nodes and nine with eight. Pr\"ufer sequences connected each set of $N$ nodes with $N-1$ initial struts; intersecting struts were joined by inserting nodes at their crossings. Sampling 250 coordinate sets and 250 connections per arrangement produced 812,500 candidate graphs. We scaled each graph to a $10~\mathrm{mm}$ cubic cell and assigned a common strut radius to maintain a relative density of $0.05$. For the $d_{31}$--$d_{33}$ comparison, we used 705 mechanically admissible Materials Project tensors and 19 bulk-ceramic entries with both coefficients reported in the literature. (Supplementary Section~1). Before calculating the piezoelectric response, we imposed a voltage of $20~\mathrm{kV}$ between the opposite $z$ faces, equivalent to a nominal field of $2~\mathrm{MV\,m^{-1}}$ to simulate poling. We solved the potential at each internal junction using the conductance-network approximation
\begin{equation}
\sum_{m\in\mathcal N(n)}g_{nm}(\phi_n-\phi_m)=0,
\qquad g_{nm}\propto\ell_{nm}^{-1},
\label{eq:methods-network-poling}
\end{equation}
where $\mathcal N(n)$ contains the nodes connected to $n$, $\phi_n$ is its potential and $\ell_{nm}$ is the connecting strut length. For a strut joining nodes $u$ and $v$, the local field is $\bm E_b=-(\phi_v-\phi_u)\bm t_b/\ell_b$, where $\bm t_b$ is its unit direction. The field direction sets the local polarization axis and its magnitude scales the axial piezoelectric coupling according to the field-dependent poling relation in Supplementary Section~2.3. Struts with zero local field were treated as unpoled.

\subsection*{Static beam model and tensor homogenization}

We computed the effective response in Abaqus/Standard with a coupled, two-node Timoshenko-beam user element, using 20 elements per strut. (Supplementary Fig.~S2) Each node carries three displacements, three rotations and an electric potential. The beam can stretch, bend, shear and twist mechanically, while its direct electrical coupling acts through axial strain. For strut $b$, the local force--charge relation is
\begin{equation}
\begin{aligned}
\bm n_b&=\bm C_b\bm g_b-\bm\Theta_b E_b,\qquad
q_b=\bm\Theta_b^{\mathsf T}\bm g_b+\epsilon_s A_b E_b,\\
\bm\Theta_b&=A_b e_{\mathrm{ax},b}(1,0,0,0,0,0)^{\mathsf T}.
\end{aligned}
\label{eq:methods-beam-law}
\end{equation}
where $\bm g_b$ contains axial strain, bending curvatures, transverse shear strains and twist; $\bm n_b$ contains their corresponding force and moment resultants; $A_b$ is the cross-sectional area; and $q_b$ is the electric-displacement resultant. The poling field sets the signed axial coupling $e_{\mathrm{ax},b}$. The fully poled value was $e_{33}^{\mathrm{sat}}=4.464~\mathrm{C\,m^{-2}}$; the reduced beam model used $E=10~\mathrm{GPa}$, $\nu=0.35$, density $7500~\mathrm{kg\,m^{-3}}$ and permittivity $\epsilon_s=2.6839\times10^{-8}~\mathrm{F\,m^{-1}}$ (Supplementary Section~2.2). Connected struts convert their local axial actuation into normal and shear motion of the complete cell.
We applied periodic mechanical and electrical boundary conditions and solved three static cases, each with a macroscopic field $\overline E_i$ along one Cartesian axis and zero macroscopic mechanical load. Differences in the displacements of opposite cell faces give the six averaged strains $\overline S_J$, ordered as $xx$, $yy$, $zz$, $yz$, $xz$ and $xy$, with engineering shear for the last three entries. We obtained each row of the effective tensor from
\begin{equation}
d_{iJ}^{\mathrm{eff}}=\frac{\overline S_J^{(i)}}{\overline E_i},
\qquad i=1,2,3,\quad J=1,\ldots,6.
\label{eq:methods-static-d}
\end{equation}
Thus, three field directions yield all 18 entries of the $3\times6$ tensor. We retained 611,239 complete calculations with finite outputs and coefficients below $2000~\mathrm{pC\,N^{-1}}$ in absolute value. The beam equations, poling scale and acceptance rules are given in Supplementary Section~2.

\subsection*{Graph neural network and inverse design}

We trained a graph neural network to predict the 18 static coefficients from the unit-cell geometry. Its node inputs were the three normalized coordinates; its 20 edge inputs comprised strut length, radius and the 18 components of the locally oriented PZT tensor. Each strut appeared as two directed edges. Five message-passing blocks of width 256 combined information at connected junctions, using residual updates, batch normalization, SiLU activations and dropout of 0.1. Separate attention-weighted sums pooled node and edge states; a 128-feature decoder returned the tensor. The model has 1,853,508 trainable parameters. We used 580,677 structures for training, 15,280 for validation and 15,282 for testing; training used AdamW, batch size 512, an initial learning rate of $10^{-3}$ and the checkpoint with the lowest validation error. The held-out test error was $21.44~\mathrm{pC\,N^{-1}}$ per coefficient on average (Supplementary Section~5).
For inverse design, we ranked existing graphs against the requested tensor or coefficient objective, then adjusted the free coordinates of several starting graphs while keeping their connections and periodic node pairing fixed. At each update, we recalculated strut lengths and directions, changed the common radius to preserve density, updated local piezoelectric features and differentiated the predicted tensor with respect to the free coordinates. Full-tensor matching minimized the Frobenius distance between standardized predicted and target tensors. Single-coefficient design maximized $|d_{36}|$ and the shear-selective design increased $|d_{34}|+|d_{35}|+|d_{36}|$ relative to the normal coefficients. We independently recalculated the selected designs with the beam model (Supplementary Section~6).

\subsection*{Dynamic tensor calculation}

We used the same poled beam models for steady-state harmonic calculations, adding structural mass and a loss factor $\eta=0.04$. For a field applied along direction $i$, the mechanical response after eliminating prescribed electrical potentials can be written
\begin{equation}
\left[(1+\mathrm{i}\eta)\bm K-\omega^2\bm M\right]
\widehat{\bm u}^{(i)}=\bm B_i\widehat E_i,
\qquad \omega=2\pi f,
\label{eq:methods-harmonic}
\end{equation}
where $\bm K$ and $\bm M$ are the assembled stiffness and mass matrices, and $\bm B_i$ is the piezoelectric force pattern per unit field. We extracted the complex, face-averaged strain from the displacement solution and calculated
\begin{equation}
\widehat d^{\mathrm{eff}}_{iJ}(\omega)
=\frac{\widehat{\overline S}_J^{(i)}(\omega)}{\widehat E_i(\omega)}.
\label{eq:methods-dynamic-d}
\end{equation}
We evaluated all three field directions at 123 frequencies from the quasi-static limit to $100~\mathrm{kHz}$. Independently calculated eigenmodes identified the strain carried by each mode and the work done on it by the electric-field; together these quantities explain the coefficient-specific peaks (Supplementary Section~7).

\subsection*{Printing and experimental poling}

We prepared a printable suspension from 30--40~wt\% PZT-5H powder, 10--15~mL polyethylene glycol diacrylate, 0.1--0.4~g Irgacure 819 photoinitiator and 3--5~g lead nitrate. Two selected unit cells were each repeated $3\times3\times3$ times and printed by projection stereolithography as approximately $5~\mathrm{mm}$ green bodies, using $25~\mu\mathrm{m}$ layers exposed for $1.8~\mathrm{s}$ each \cite{Cui2019,Cui2022}. We removed the binder under argon at $600~{}^\circ\mathrm{C}$ for 3~h, then sintered the lattices at $1170~{}^\circ\mathrm{C}$ for 6~h in air with a PbO-buffered PZT powder bed. The cooled ceramic lattices measured approximately $4.5~\mathrm{mm}$ on each side (Supplementary Section~8).
We placed temporary silver-epoxy electrodes against the opposite $z$ faces and immersed each ceramic lattice in silicone oil. After heating the bath to approximately $100~{}^\circ\mathrm{C}$, we increased the voltage to about $6~\mathrm{kV}$, giving an average field of $1.3$--$1.5~\mathrm{kV\,mm^{-1}}$ across the sintered lattice. We maintained the field for approximately 1~h while the sample cooled, then removed the electrodes and rinsed away residual oil. This step aligns ferroelectric domains in the printed PZT struts (Supplementary Section~8).

\subsection*{Piezoelectric characterization}

We measured the direct coefficient using a Berlincourt meter. The meter applies a small oscillating compressive force along the poling direction and measures the charge generated at the electrodes on the opposite $z$ faces. The ratio of charge to force gives $d_{33}=Q_3/F_3$, expressed in $\mathrm{pC\,N^{-1}}$.
We measured the converse response using a laser Doppler vibrometer (PSV-500, Polytec) under a voltage applied across the $z$ faces. The instrument detects the Doppler shift of light reflected from the moving surface to determine its velocity, from which displacement is obtained. A lightweight L-shaped reflector provided orthogonal surfaces for measuring vertical and lateral motion. We calculated the longitudinal response as $d_{33}=\Delta u_z/V$ and reported $\Delta u_x/V$ as an apparent $d_{35}$ under the stated mounting conditions, both in $\mathrm{pm\,V^{-1}}$. The measured longitudinal coefficients were consistent between the two methods (Supplementary Section~9.1).

\section*{Author contributions}

Conceptualization AS, MM and XRZ. Data collection: AS, MM. HC Simulation analysis: AS, MM. HC. Device fabrication: NM and VC. Testing: AS, VC. Supervision: XRZ, MM. Writing-original draft: AS, MM, XRZ. Writing-review and editing: AS, MM, XRZ

\section*{Ethics declarations}

\textbf{Competing interests.} The authors declare no competing interests.

\section*{Supplementary information}

Data will be provided upon publication of the work

\section*{Data availability}

Data will be provided upon publication of the work

\bibliographystyle{unsrtnat}
\bibliography{references}

\clearpage

\clearpage

\clearpage

\clearpage

\clearpage

\end{document}